\documentclass{elsart}
\usepackage{graphics,epsfig,amsmath,amssymb,amstext}

\def\d {\mathrm{d}}
\def\pt {p_{\perp}}

\def\degree{\hbox{$^\circ$}}

\begin{document}

    \begin{frontmatter}
	
	\title{Multiple UHECR Events from Galactic hadron jets}

	\author{Etienne Parizot}

	\address{Institut de Physique Nucl\'eaire d'Orsay,
	IN2P3-CNRS/Universit\'e Paris-Sud, 91406 Orsay Cedex, France}	

	\begin{abstract}
	    We propose a new observational test of top-down source
	    models for the ultra-high-energy cosmic-rays (UHECRs),
	    based on the simultaneous observation of two or more
	    photons from the same Galactic hadron jet.  We derive a
	    general formula allowing one to calculate the probability
	    of detecting such `multiple events', for any particular
	    top-down model, once the physical parameters of the
	    associated hadron jets are known.  We then apply our
	    results to a generic top-down model involving the decay of
	    a supermassive particle, and show that under reasonable
	    assumptions the next-generation UHECR detectors would be
	    able to detect multiple events on a timescale of a few
	    years, depending on the mass of the top-down progenitor. 
	    Either the observation or the non-observation of such
	    events will provide constraints on the UHECR top-down
	    models and/or the physics of hadronization at ultra-high
	    energy.
	\end{abstract}

    \end{frontmatter}
    
    \section{Introduction}
    
    Ultra-high-energy cosmic rays (UHECRs) are puzzling in respect of
    both their production and their propagation in the universe.  On
    the one hand, even the most powerful astrophysical sites known to
    be able to accelerate particles to very high energy seem to have
    difficulties to reach energies as high as $3\times 10^{20}$~eV
    (the highest reported UHECR energy so far\cite{HighestEnergies}). 
    On the other hand, even if they could, one would expect from the
    (presumably) extremely high rigidity of the observed UHECRs in the
    intergalactic medium that their directions of arrival in the Earth
    atmosphere roughly point towards the sources, which does not seem
    to be the case.  Also, it had been expected that the UHECR flux
    above $\sim 10^{20}$~eV would be very much reduced due to the
    interaction of the UHE particles with the cosmological microwave
    background.  This so-called GZK cutoff\cite{GZKCutoff}, however,
    does not seem to be present in the currently available data.
    
    Although the UHECR sources are still essentially unknown, many
    models have been proposed, with various charms and problems
    \cite{UHECRReview}.  They can be divided up into two classes:
    \textit{bottom-up} models, in which particles initially at low
    (thermal) energy get accelerated by one or a series of
    astrophysical processes, and \textit{top-down} models, in which
    each UHECR is directly produced, \textit{as a particle}, at
    ultra-high energy, through the decay of a pre-existing
    supermassive particle or some exotic, high-energy physical
    process, e.g. involving the collapse or annihilation of
    topological defects.  In this paper, we consider the so-called
    Galactic top-down models from a general point of view, merely
    assuming that the UHECR flux is dominated by sources in the Halo,
    with a density proportional to that of the dark matter.  This
    solves the `production problem' trivially (or more exactly shifts
    it to the problem of identifying the X-particles and explaining
    their production and decay rates) and provides a simple
    understanding of the absence of a GZK cutoff as well as the
    apparent isotropy of the UHECRs -- at least until the statistics
    will be high enough for us to detect the dipole anisotropy due to
    the off-centered position of the solar system in the Galaxy (which
    should take at least three years of observation with the Pierre
    Auger Observatory (PAO)~\cite{MedWat99}).
    
    Many different models have been proposed, with X-particles of
    different types (either produced locally, notably through
    topological defect interactions, or inherited from the big bang)
    and different masses (on the Planck scale, $10^{28}$~eV, the GUT
    scale, $10^{25}$~eV, or below)~\cite{UHECRReview,TopDownModels}. 
    In this paper, we investigate a common consequence of a large
    class of Galactic top-down models, and propose an observational
    test which could be accessible to the next generation of UHECR
    detectors, such as the PAO~\cite{PAO}, the EUSO
    experiment~\cite{EUSO} or the OWL/AirWatch project~\cite{OWL}.
    
    \section{Multiple UHECR events: the basic idea}
    \label{sec:basicIdea}
    
    Top-down scenarios involve the production of \textit{hadron jets}
    in a way similar to what is observed in terrestrial accelerators,
    when an energetic quark-antiquark pair (e.g. produced through
    e$^{+}$e$^{-}$ annihilation) \textit{hadronizes} into a number of
    colourless hadrons through a QCD cascade.  Let $N_{\gamma}$ be the
    number of photons in a jet (from neutral pion decay), and
    $\theta_{\mathrm{jet}}$ be the jet opening angle.  Since
    gamma-rays propagate in straight lines away from the source, the
    average surface density of UHE photons at a distance $D$ from the
    point where the X-particle decayed is $\sigma =
    N_{\gamma}/\omega_{\mathrm{jet}}D^{2}$, where
    $\omega_{\mathrm{jet}} \simeq \pi\theta_{\mathrm{jet}}^{2}$ is the
    jet solid angle.  If a detector intersects such a jet, with a
    surface area $S_{\perp}$ orthogonal to the jet axis, it will see
    on average the following number of particles:
    \begin{equation}
	\mu =
	\frac{N_{\gamma}S_{\perp}}{\pi\theta_{\mathrm{jet}}^{2}D^{2}}.
	\label{eq:multiplicity}
    \end{equation}    
    If the source is close enough, the number of photons in the jet
    high enough and the detector surface large enough, then $\mu$ will
    be larger than one and several UHECRs will be able to cross the
    detector at (almost exactly) the same time, from (almost exactly)
    the same direction.  This is what we define to be a
    \textit{multiple event}.  It has an unambiguous experimental
    signature: two or more distinct showers developing simultaneously
    in the atmosphere, with almost perfectly parallel axes (within
    $\sqrt{S_{\perp}}/D$ radians, which is much less than any
    conceivable experimental angular resolution).\footnote{Note that
    this is very different from the clustered events, sometimes
    referred to in the literature as \textit{multiplets}, which
    correspond to independent UHECRs arriving from roughly the same
    direction in the sky, but at different times.}
    
    The number $\mu$ may be called the multiplicity of the X-particle
    decay event, or more exactly its \textit{potential multiplicity},
    as can be expected at Earth, since it is the average number of
    particles which can be observed simultaneously by the detector
    (assuming that it intersects the jet).  For a given X-particle
    decay event, with a given $\mu$, the \textit{actual multiplicity}
    of the UHECR event as observed by the detector can only be
    predicted statistically.  The probability, $\mathcal{P}(m,\mu)$,
    of observing an event with actual multiplicity $m$ (integer) in a
    jet of potential multiplicity $\mu$ (real number) is given by the
    binomial law:
    \begin{equation}
	\mathcal{P}(m,\mu) =
	C_{N_{\gamma}-1}^{m-1}\left(\frac{\mu}{N_{\gamma}}\right)^{m-1}
	\left(1-\frac{\mu}{N_{\gamma}}\right)^{N_{\gamma}-m},
	\label{eq:probaOfMulti}
    \end{equation}
    where $\mu/N_{\gamma} = S_{\perp}/S_{\mathrm{jet}}$ is the ratio
    of the detector's surface to the jet surface (see
    Eq.~(\ref{eq:multiplicity})), and thus the probability for a given
    particle in the jet to cross the detector.  Note that
    $\mathcal{P}(m,\mu)$ is actually the conditional probability of
    the multiple event, given the fact that one shower is observed, or
    if one prefers, the probability that a detected shower be
    accompanied by $m-1$ others.  The probability of observing a
    multiple event with whatever multiplicity larger than two simply
    adds up to $\mathcal{P}(m\ge 2,\mu) = 1 - (1 -
    \mu/N_{\gamma})^{N_{\gamma}-1} \approx 1 - e^{-\mu}$, for not too
    small values of $N_{\gamma}$.
    
    The basic idea behind top-down multiple events is thus that the
    UHECRs are not independent of one another, but appear in close
    groups released at the same time in a single X-particle decay
    event.  If the groups are sufficiently tight, we should be able to
    detect several UHECRs at a time.  In fact, top-down jets can be
    seen as genuine \textit{Galactic showers}: in the same way as we
    detect single UHECR events by intercepting many secondary
    particles belonging to the same \textit{atmospheric} shower, the
    use of very large detectors may allow us to detect single
    `X-particle decay events' by intercepting several UHECRs belonging
    to the same \textit{Galactic} shower.  The detectability of
    multiple events thus comes down to the question: when we see one
    UHECR in a top-down jet, how close is the next one, compared to
    the detector's radius?

    \section{Timescale of multiple event detection}
    
    The potential multiplicity, $\mu$, of a UHECR event
    (Eq.~\ref{eq:multiplicity}) depends on two physical parameters
    related to the jet properties, $N_{\gamma}$ and
    $\theta_{\mathrm{jet}}$, one astrophysical parameter, $D$, related
    to the source distribution, and one `experimental' parameter,
    $S_{\perp}$, related to the detector.  Most X-particle decays will
    occur much too far from the solar system to give rise to multiple
    events.  But if one assumes that the X-particles distribute over
    the Galactic halo in the same way as the dark-matter, one can
    estimate the probability that one of the many UHECR events that
    will be detected over a given period of observation corresponds to
    a small enough source distance.
    
    \begin{figure}
	\centering
	\includegraphics[width=9cm]{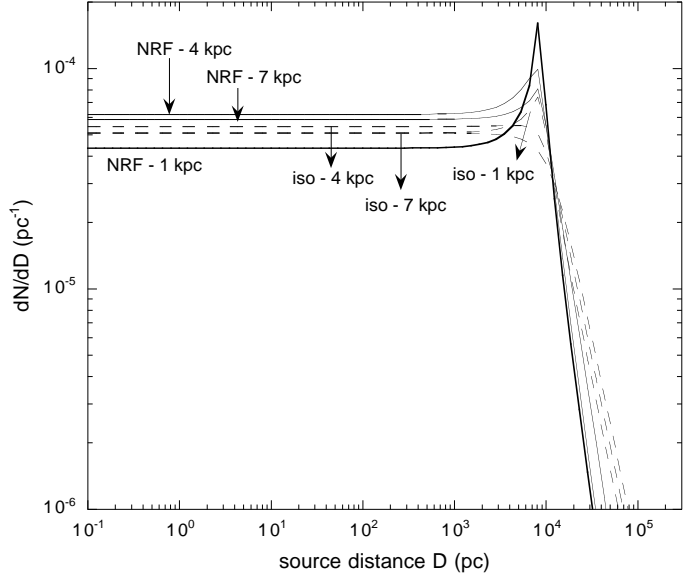}
	\caption{Effective source density for various models of the
	dark-matter distribution in our Galaxy.  Labels indicate the
	assumed value of $R_{\mathrm{c}}$ for either isothermal or FRW
	models (see text).}
	\label{fig:sourceDensity}
    \end{figure}
    
    \subsection{The distribution of source distances}
    
    The statistics of multiple events depend on that of source
    distances.  For the dark-matter distribution in the Galaxy, we may
    consider either a simple isothermal halo model \cite{CalOst81},
    where the density, $n_{\mathrm{DM}}$ depends on the galactocentric
    distance, $r$, proportionally to $1/(r^{2} + R_{\mathrm{c}}^{2})$,
    and the core radius $R_{\mathrm{c}}$ is of the order of a few
    kiloparsecs, or an FRW model based on cold dark matter
    simulations, with $n_{\mathrm{DM}}\propto
    1/[r(r+R_{\mathrm{c}})^{2}]$ \cite{Navarro+96}.  In
    Fig.~\ref{fig:sourceDensity}, we have plotted the corresponding
    effective UHECR source density as a function of distance, for an
    observer located at the galactocentric radius of the Sun (taking
    into account the smaller effective contribution of more distant
    sources).
    
    As can be seen, the effective source density is flat for low
    values of the source distance, which are those of interest to us
    because they give the highest probability of observing multiple
    events.  This result is nothing but the famous Olber's paradox,
    and it is independent of the actual dark matter distribution,
    provided it is not varying significantly on small scales. 
    Moreover, the actual density profile of the dark matter halo
    appears not to affect significantly the normalization of the
    source density at small distances (except for unreasonable values
    of $R_{\mathrm{c}}$).  In the following, we adopt the value of
    $6\times 10^{-5}\,\mathrm{pc}^{-1}$ (recalling that NRF models are
    currently preferred) and replace, for all practical purposes of
    the present study, the effective distribution of UHECR source
    distances by the following differential probability:
    \begin{equation}
	    \d\mathrm{P}(D) = p(D)\d D = \frac{1}{D_{0}} \d D
	    \quad\quad (0\le D\le D_{0}),
	\label{eq:p(D)}
    \end{equation}
    where $D_{0} = 1/(6\times 10^{-5}) \simeq 17$~kpc is an effective
    radius beyond which no UHECR sources exist (that is, they
    contribute a negligible flux at Earth).
    
    Note that inhomogeneities in the dark matter distribution may in
    practice alter the probability of X-particle decay events at a
    given point of the Galaxy.  A lower concentration of sources close
    to the Earth would decrease the chance of detecting multiple
    events, while a higher concentration would increase it.  Lacking a
    precise knowledge of the small scale dark matter distribution, we
    can but assume that the Earth environment is not very different
    from the average.
    
    \subsection{Multiple event probability}
    
    For convenience, we shall rewrite the potential multiplicity of an
    individual UHECR event, given by Eq.~(\ref{eq:multiplicity}), as:
    \begin{equation}
	\mu(D) = \mu_{0}\frac{D_{0}^{2}}{D^{2}},\quad\mathrm{where}
	\quad\mu_{0} = \frac{N_{\gamma}S_{\perp}}
	{\pi\theta_{\mathrm{jet}}^{2}D_{0}^{2}}.
	\label{eq:muMu0}
    \end{equation}
    For any given model, the probability of detecting an event of
    actual multiplicity larger than $n$ increases with the total
    number of UHECR events detected, $N_{\mathrm{evt}}$, according to
    the simple law:
    \begin{equation}
	\mathcal{P}_{\ge n}(N_{\mathrm{evt}}) = 1 -
	e^{-\frac{N_{\mathrm{evt}}}{N_{n}}},
        \label{eq:globalProbaMultiN}
    \end{equation}
    where the constants $N_{n}$ are the characteristic numbers of
    events which have to be detected before it becomes reasonably
    probable ($\sim 63\%$) to detect an event of multiplicity $n$. 
    This is a straightforward consequence of the statistical
    independence of X-particle decay events (see Appendix).
    
    We show in the appendix that the characteristic event number for
    double event detection is given by~:
    \begin{equation}
	N_{2} = \frac{1}{\sqrt\pi}\mu_{0}^{-1/2},
	\label{eq:N2}
    \end{equation}
    and that the subsequent numbers deduce from $N_{2}$ by the
    following recursion relation (valid for detected multiplicities
    much smaller than the total jet multiplicity, $n\ll N_{\gamma}$):
    \begin{equation}
        N_{n+2} = \frac{2n}{2n-1}N_{n+1},
        \label{eq:recursionOnN}
    \end{equation}
    so that, in particular, $N_{3} = 2N_{2}$, $N_{4} =
    \frac{4}{3}N_{3}$, $N_{5} = \frac{6}{5}N_{3}$, etc.
    
    \subsection{Multiple event detection timescales}
    
    In order to convert the above characteristic event numbers into
    multiple event detection timescales (for a given detector), we
    just need to calculate the UHECR detection rate.  This depends on
    the total aperture, $\mathcal{A}_{\mathrm{d}}$ (in
    $\mathrm{km}^{2}\mathrm{sr}$), and the duty cycle, $\delta$ (in
    percent), of the detector.  The UHECR detection rate above energy
    $E$ is given by:
    \begin{equation}
	\dot{N}_{\mathrm{evt}}(\ge E) = \Phi_{\mathrm{CR}}(\ge
	E)\times \mathcal{A}_{\mathrm{d}}\times\delta\,,
        \label{eq:NDot}
    \end{equation}
    where $\Phi_{\mathrm{CR}}(\ge E)$ is the integral flux of UHECRs
    above energy $E$.  From the AGASA and Fly's Eye experiments, a
    fair value of the UHECR flux at $10^{20}$~eV is
    $\Phi_{\mathrm{CR}}(10^{20}) \simeq 3\times 10^{-40}\,
    \mathrm{cm}^{-2}\mathrm{s}^{-1}\mathrm{sr}^{-1}\mathrm{eV}^{-1}$. 
    The intergal flux obviously depends on the actual spectrum, which
    is virtually unknown above $10^{20}$~eV. As for the spectrum below
    that energy, one should keep in mind that we are only interested
    in the events which can be attributed to the top-down process
    under consideration, and which probably represent only a fraction
    of the total detected events between $10^{19}$ and $10^{20}$~eV.
    We shall note $\Phi^{\mathrm{td}}_{\ge E_{\mathrm{th}}}$ the
    corresponding integrated flux above the detector's threshold
    energy, $E_{\mathrm{th}}$.
    
    We can now express the time evolution of the multiple event
    probabilities, by replacing $N_{\mathrm{evt}}$ by
    $\dot{N}_{\mathrm{evt}}\times t$ in
    Eq.~(\ref{eq:globalProbaMultiN}):
    \begin{equation}
	\mathcal{P}_{\ge n}(t) = 1 - e^{-\frac{t}{\tau_{n}}},\quad
	\mathrm{where}\quad \tau_{n} = N_{n}/\dot{N}_{\mathrm{evt}}.
	\label{eq:globalProbaMultiP}
    \end{equation}
    Using Eqs.~(\ref{eq:N2}) and~(\ref{eq:NDot}) and the expression
    for $\mu_{0}$, Eq.~(\ref{eq:muMu0}), we find:
    \begin{equation}
	\tau_{2} =
	\frac{\theta_{\mathrm{jet}}(E_{\mathrm{th}})\,D_{0}}
	{N_{\gamma,\ge E_{\mathrm{th}}}^{1/2}
	\left<S_{\perp}\right>^{1/2} \Phi^{\mathrm{td}}_{\ge
	E_{\mathrm{th}}} \mathcal{A}_{\mathrm{d}}\delta}\,\,,
	\quad\mathrm{and}\quad\tau_{n+2} = \frac{2n}{2n-1}\tau_{n+1}.
	\label{eq:tau2Compact}
    \end{equation}
    
    In practice, the average perpendicular surface of the detector,
    $\left<S_{\perp}\right>$, is related to the acceptance
    $\mathcal{A}_{\mathrm{d}}$.  If the detector's surface \textit{on
    the ground} is $S_{\mathrm{d}}$, and $\theta_{\mathrm{max}}$ is
    the maximum zenith angle visible by the detector, we have
    $\left<S_{\perp}\right> =
    \frac{1}{2}S_{\mathrm{d}}\times\sin^{2}\theta_{\mathrm{max}}/(1 -
    \cos\theta_{\mathrm{max}})$, and $\mathcal{A}_{\mathrm{d}} =
    S_{\mathrm{d}}\times\pi\sin^{2}\theta_{\mathrm{max}}$.  If
    $\theta_{\mathrm{max}} = 90\degree$, $\left<S_{\perp}\right> =
    \tfrac{1}{2}S_{\mathrm{d}}$, and $\mathcal{A}_{\mathrm{d}} = \pi
    S_{\mathrm{d}}$.  Reporting into Eq.~(\ref{eq:tau2Compact}), we
    get:
    \begin{equation}
	\tau_{2} = \frac{\sqrt
	2\,\theta_{\mathrm{jet}}(E_{\mathrm{th}}) \,D_{0}}
	{\pi\,N_{\gamma,\ge E_{\mathrm{th}}}^{1/2}
	\Phi^{\mathrm{td}}_{\ge E_{\mathrm{th}}}
	S_{\mathrm{d}}^{3/2}\delta},
	\label{eq:tau2CompactBis}
    \end{equation}

    \section{Numerical estimates for a toy jet model}
    \label{sec:jetModel}
    
    The remaining parameters necessary to calculate $\tau_{2}$ are the
    jet parameters, namely the photon multiplicity in the jet,
    $N_{\gamma}$, or more precisely the number of photons as a
    function of energy, $(\mathrm{d}N_{\gamma}/\mathrm{d}E)(E)$, and
    the jet opening angle, $\theta_{\mathrm{jet}}$.  Unfortunately,
    they are the most uncertain, because the detailed structure of the
    hadron jets produced at the ultra-high energies of interest is not
    known, and one can only extrapolate from the semi-empirical models
    available at CERN energies, assuming that nothing dramatic occurs
    in physics at the intermediate energy scales.  We shall not
    attempt here to describe QCD jet physics and theory, and refer the
    reader to the review by Bhattachargee and Sigl~\cite{UHECRReview}
    of the various models, and to the book of Dokshitzer et
    al.~\cite{Dokshitzer+91}, notably chapters 7 and 9, where the
    energy spectrum and multiplicity of the particles in a jet are
    discussed in detail, as well the collimation of both particles and
    energy.  An interesting discussion of UHECR spectra in top-down
    models can also be found in~\cite{Sarkar+00}.
    
    \subsection{The photon multiplicity in a jet}
    
    \begin{figure}
       \centering
       \includegraphics[width=6.6cm]{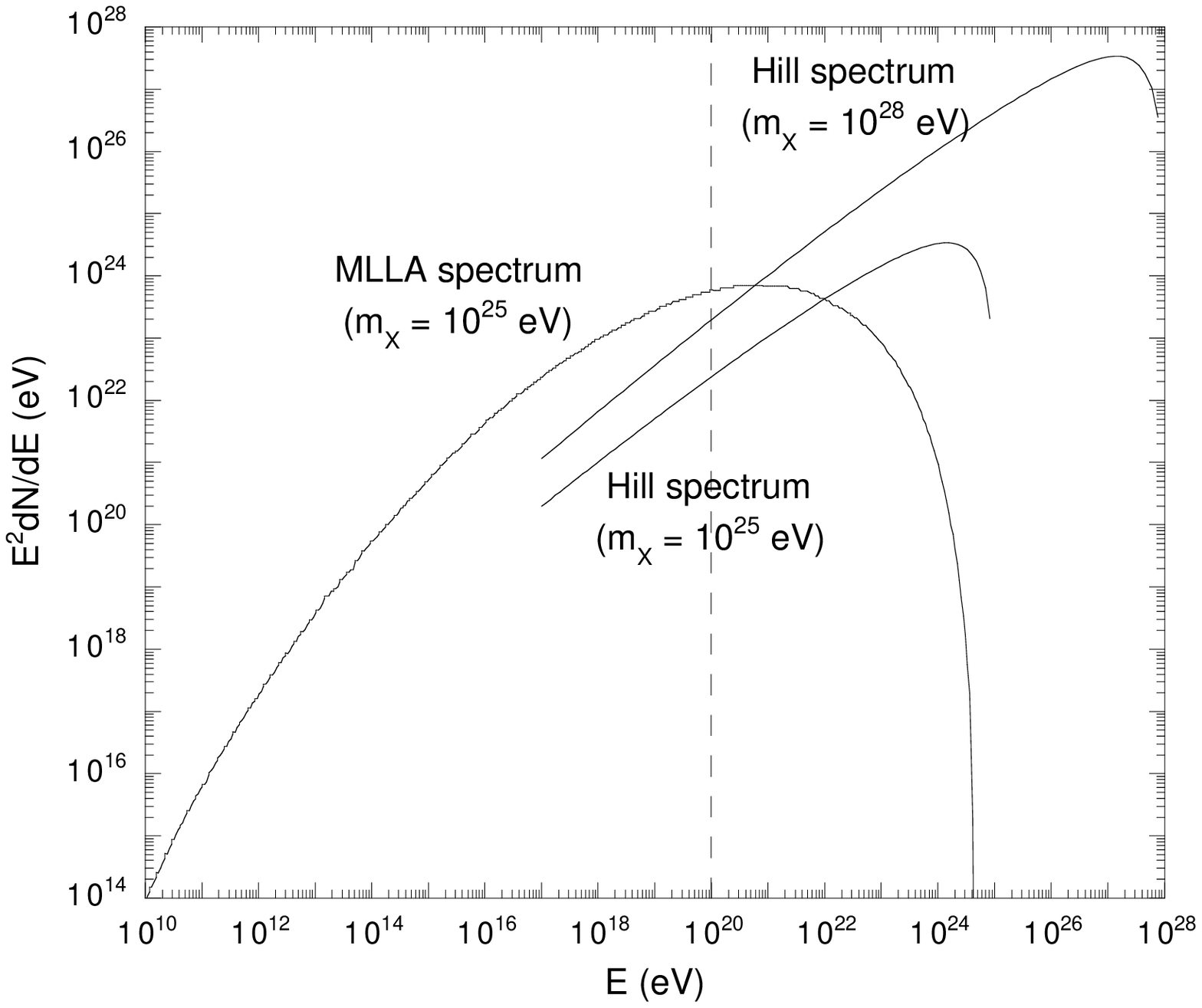}\hspace{0.1cm}
       \includegraphics[width=6.6cm]{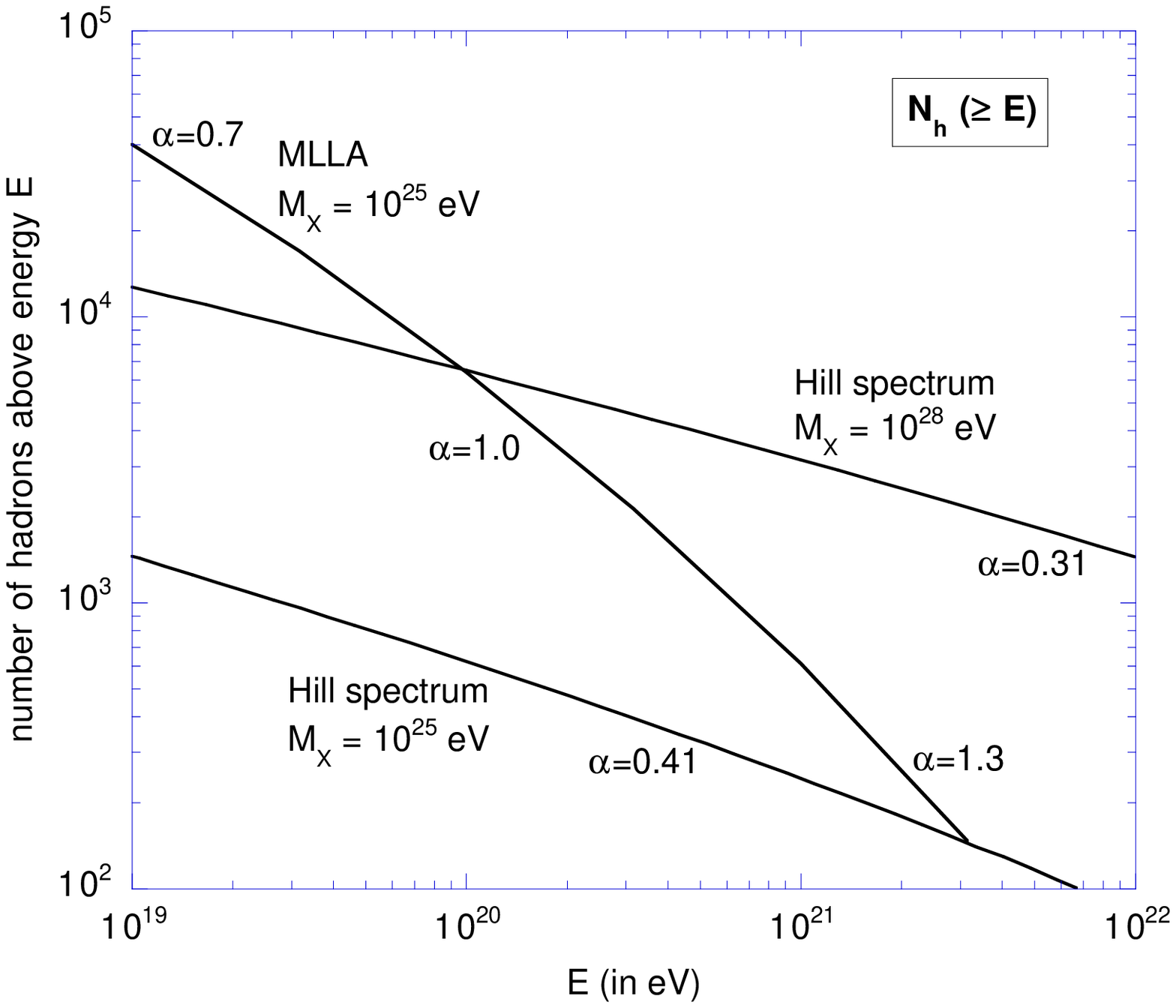}
       \caption{Left: Differential energy spectrum (multiplied by
       $E^{2}$) of the secondary photons in a top-down jet, for
       various models (adapted from Bhattacharjee and Sigl
       \cite{UHECRReview}).  Right: Integrated energy spectrum giving
       the total number of photons above energy $E$ in the jet, for
       the same models.  The slope of the approximate power-law
       (spectral index $\alpha$) is indicated.}
       \label{fig:hadronJetSpectrum}
    \end{figure}

    Concerning the jet particle multiplicity, we show on
    Fig.~\ref{fig:hadronJetSpectrum}a the typical spectrum (multiplied
    by $E^{2}$) obtained with a modified leading-log approximation
    (MLLA) model, assuming that the X-particle mass is at the GUT
    scale, i.e. $M_{\mathrm{X}} \simeq 10^{25}~\mathrm{eV/c}^{2}$
    (adapted from~\cite{UHECRReview}).  Also shown are the hadron jet
    spectra obtained with the Hill formula for X-particle masses of
    $10^{25}~\mathrm{eV/c}^{2}$ and $10^{28}~\mathrm{eV/c}^{2}$.  In
    all cases, it is expected that most of the jet energy be
    distributed among UHE particles.  Following some previous works,
    we consider here MLLA spectra, which are found not to have a
    simple power-law behavior in the energy range of interest, namely
    around $10^{20}$~eV, and to be steeper than often quoted for
    top-down scenarios -- as would result from Hill's formula.  It has
    been argued, however, that the MLLA spectra do not reproduce
    faithfully the fragmentation spectrum in the last energy decade or
    so, i.e. at energies close to the X-particle
    mass~\cite{Sarkar+00}.  In this study, we shall only consider
    X-particle masses above $10^{23}$~eV, so that the UHECRs of
    interest have energies well below $m_{\mathrm{X}}$.  The MLLA
    spectral shape may thus represent a reasonable approximation of
    the fragmentation spectrum around $10^{20}$~eV. The normalization
    is probably more problematic, however, because it depends largely
    on the amount of energy carried out by the rarest, most energetic
    particles in the jet.  Any other normalization than adopted below
    will lead to characteristic timescales for multiple event
    detection that scale according to Eq.~(\ref{eq:tau2CompactBis}),
    as $N_{\gamma}^{-1/2}$.
    
    In Fig.~\ref{fig:hadronJetSpectrum}b, we have plotted the
    integrated hadron jet spectrum corresponding to the same cases as
    in Fig.\ref{fig:hadronJetSpectrum}a.  If one tries to approximate
    the spectrum by a power-law $\d N/\d E\propto E^{-x}$ (i.e. $N(\ge
    E) \propto E^{-\alpha}$, with $\alpha = x - 1$), the logarithmic
    slope is approximately constant for the Hill spectra, while it
    goes from 1.7 to 2.3 in the energy range of interest, with a value
    of $\sim 2$ at $10^{20}$~eV, for the MLLA spectrum.  In our `toy
    jet model', we shall assume a mean hadron spectrum in $E^{-2}$
    between $E_{\mathrm{inf}} = 10^{19}$~eV and $E_{\mathrm{sup}} =
    10^{22}$~eV: $N(E) = \eta E_{\mathrm{jet}}/E^{2}$, where $\eta$ is
    a numerical constant, so that $N(\ge E) = \eta(E_{\mathrm{jet}}/E
    - 1) \simeq \eta E_{\mathrm{jet}}/E$.  A fit of the MLLA spectrum
    in Fig.~\ref{fig:hadronJetSpectrum} gives $\eta\simeq 0.10$, so
    that about 70\% of the jet energy is in particles with energies
    above $10^{19}$~eV. Only about one third of this energy, however,
    will be imparted to the photons, assuming that the total jet
    energy is divided up evenly into the three types of pions
    ($\pi^{0}$, $\pi^{+}$ and $\pi^{-}$), of which only the neutral
    ones decay into photons, and neglecting in a first approximation
    the contribution of nucleons.
    
    In conclusion, we shall adopt for the UHE photons the above
    $E^{-2}$ spectrum with a value of $\eta = 0.033$.  We shall also
    extrapolate the same spectrum to hadron jets generated by
    X-particles of lower mass, but with $E_{\mathrm{jet}}$,
    $E_{\mathrm{inf}}$ and $E_{\mathrm{sup}}$ scaled linearly.  This
    will allow us to explore UHECR progenitors with masses down to
    $10^{23}$~eV. Assuming that the X particle decay events lead to
    the formation of two jets, so that $E_{\mathrm{jet}} =
    M_{\mathrm{X}}c^{2}/2$, one finally obtains the following
    approximate formula (only valid between $E_{\mathrm{inf}}$ and
    $E_{\mathrm{sup}}$):
    \begin{equation}
	N_{\gamma}(\ge E) \simeq 1.7\times 10^{3}
	\left(\frac{E}{10^{20}\,\mathrm{eV}}\right)^{-1}
	\left(\frac{M_{\mathrm{X}}c^{2}}{10^{25}\,\mathrm{eV}}\right).
        \label{eq:jetMultiplicity}
    \end{equation}
    
    \subsection{The jet opening angle}
    
    Coming now to the question of the jet opening angle, let us first
    note that a naive line of reasoning based on the Lorentz factor
    collimation effect cannot apply here.  One might have been tempted
    to derive $\theta_{\mathrm{jet}}$ by claiming that an isotropic
    distribution of the jet particles in the rest frame of the parent
    quark would translate in the Galactic frame into a collimated
    distribution within a cone of opening angle $1/\Gamma_{q}$, where
    $\Gamma_{q}$ is the quark's Lorentz factor.  However, such a
    collimation only applies to the decay products of \textit{real}
    particles, not \textit{virtual} ones.  In the case considered
    here, the parent X-particle \textit{is} at rest in the Galactic
    frame, and the jet particles are created out of the extremely
    intense field represented by a quark/anti-quark pair moving apart,
    not by the `decay' of one of its members.  In a QCD jet, as it
    turns out, the hadronization process allows in principle large
    emission angles, i.e. large values of the particle momentum in a
    direction perpendicular to the jet axis, $p_{\perp}$.
    
    A standard angular distribution is given by the following
    logarithmic law:
    \begin{equation}
	\frac{\d N}{\d\pt^{2}} \simeq \frac{1}{\pt^{2} +
	\Lambda_{QCD}^{2}},
        \label{eq:logDistrib}
    \end{equation}
    where the regularisation momentum, $\Lambda_{QCD}$, is a typical
    effective QCD scale, of the order of 300~MeV. Using this
    expression, and the fact that the emission angle of a particle is
    given by $\theta(\pt) \simeq \pt/p_{\parallel}\simeq\pt c/E$, one
    finds that about 10\% of the jet particles are found with $\pt <
    30 \Lambda_{QCD}$, i.e. within $\theta_{\mathrm{jet}}\simeq
    9\times 10^{-11}(E/10^{20}\,\mathrm{eV})^{-1}$, with only a weak
    dependence on the X-particle mass (logarithmic in the large
    quantity $M_{\mathrm{X}}c^{2}/\Lambda_{\mathrm{QCD}}$).  This
    represents a considerably weaker collimation than what would have
    been obtained from a Lorentz factor argument applied to a
    progenitor \textit{real} quark, giving the opening angle
    $\theta_{\mathrm{jet}}\simeq 1/\Gamma_{\mathrm{q}}\simeq 3\times
    10^{-14}(M_{\mathrm{X}}c^{2}/10^{25}\,\mathrm{eV})^{-1}$.
    
    In practice, however, the energy is found to be better collimated
    than the multiplicity in QCD jets~\cite{Dokshitzer+91}, and one
    expects the highest energy particles (which we are interested in)
    to be much better collimated than obtained from
    Eq.~(\ref{eq:logDistrib}).  In other words, the largest
    perpendicular momenta in the jet distribution are attributed
    statistically more often to the lower-energy particles.  Moreover,
    this behaviour is found to be amplified as the jet energy
    increases.  This is important for our concern, because we are
    interested only in the particles in the last few decades of the
    energy range (above a few $10^{19}$~eV), and not in all the (much
    more numerous) particles between this energy and the GeV range,
    which may fill a cone with a larger opening angle, but which shall
    not be detected as UHECRs anyway.
    
    Extrapolating the semi-empirical theory available at CERN energies
    towards the ultra-high energies of interest, one finds that for a
    quark jet, about 50\% of the jet energy is found within
    $\theta_{\mathrm{UHE}} \sim 2\times 10^{-12}$ radians of the jet
    axis (Dokshitzer, private communication).  Considering this number
    as well as that obtained from Eq.~(\ref{eq:logDistrib}), we shall
    arbitrarily adopt the following, hopefully conservative value for
    the UHECR jet opening angle:
    \begin{equation}
	\theta_{\mathrm{jet}} \simeq 2\times 10^{-11}.
        \label{eq:thetaJet}
    \end{equation}
    
    The above estimates are admittedly arguable, and cannot be
    expected to hold for all models, but they may represent a
    reasonable description of the jets in the energy range of
    interest.  Any other assumptions about $\d N_{\mathrm{gamma}}/\d
    E$ and $\theta_{\mathrm{jet}}$ (e.g. motivated by a detailed study
    of a particular top-down model) can be used in the following, in a
    straightforward replacement of ours.
        
    \subsection{Observability of multiple events with the PAO and EUSO}
    
    Let us now evaluate the characteristic timescale of double event
    observation, $\tau_{2}$, by replacing the various model parameters
    by their numerical values in Eq.~(\ref{eq:tau2CompactBis}).  The
    flux of top-down UHECRs above the detector's threshold energy,
    $\Phi^{\mathrm{td}}_{\ge E_{\mathrm{th}}}$, is obtained
    consistently with the assumed hadronization spectrum: we normalize
    the $E^{-2}$ spectrum to the quoted value of $3\times 10^{-40}\,
    \mathrm{cm}^{-2}\mathrm{s}^{-1}\mathrm{sr}^{-1}\mathrm{eV}^{-1}$
    at $10^{20}$~eV, i.e. we assume that all the UHECRs at
    $10^{20}$~eV have a top-down origin (and only a fraction of them
    below that energy).  One thus obtains:
    \begin{equation}
	\Phi_{\mathrm{CR}}(\ge E) \simeq (10^{-2}\,
	\mathrm{km}^{-2}\mathrm{yr}^{-1}\mathrm{sr}^{-1})\times
	\left(\frac{E}{10^{20}\,\mathrm{ev}}\right)^{-1},
    \end{equation}
    from which it follows:
    \begin{equation}
	\tau_{2} \simeq (2.1\,\,\mathrm{yr})\times
	\left(\frac{S_{\mathrm{d}}}{3000\,\mathrm{km}^{2}}\right)^{-3/2}
	\left(\frac{\delta}{100\%}\right)^{-1}
	\left(\frac{E_{\mathrm{th}}}{10^{19}\,\mathrm{eV}}\right)^{3/2}
	\left(\frac{M_{\mathrm{X}}}{10^{25}\,\mathrm{eV}}\right)^{-1/2}.
        \label{eq:tau2Developped}
    \end{equation}
    In the case of the next generation UHECR observatories, the
    detection surface on the ground will be $3000\,\mathrm{km}^{2}$
    for the PAO (one site), and $1.5\times 10^{5}\,\mathrm{km}^{2}$
    for EUSO. The detector's duty cycles are respectively 100\% and
    14\%, and the energy thresholds are $10^{19}$~eV for the PAO and
    $5\times 10^{19}$~eV for EUSO.  With these number, one finds, for
    an X-particle at the GUT scale ($M_{\mathrm{X}} =
    10^{25}\,\mathrm{eV}$):
    \begin{equation}
	\tau_{2}(\mathrm{PAO}) = 2.1\,\mathrm{yr} \quad\mathrm{and}
	\quad \tau_{2}(\mathrm{EUSO}) = 0.48\,\mathrm{yr},
    \end{equation}
    which are smaller than the observatories' lifetimes (15 and 3
    years, respectively).  The timescales for triple, quadruple and
    quintuple event detections are respectively 2 times, 2.67 and 3.2
    times larger, as follows from Eq.~(\ref{eq:tau2Compact}).
    
    As indicated above, these timescales scale with the X-particle
    mass as $M_{\mathrm{X}}^{-1/2}$, and with the actual number of UHE
    photons within the jets as $N_{\gamma}^{1/2}$.  Even a drastic
    decrease in the photon multiplicity in the jet by an order of
    magnitude would only increase the lifetimes by a factor $\sqrt 10$
    and keep the double event detection timescales smaller than the
    lifetime of each experiments.  The sensitivity to
    $\theta_{\mathrm{jet}}$ is linear, however, and jet models with
    much larger opening angles than assumed here would make double
    event detection more problematic.  We should also note that decay
    modes into more than two jets would lead to smaller photon
    multiplicities, so that $\tau_{2}$ would also scale as
    $N_{\mathrm{jet}}^{1/2}$.

    \section{Strong upper limit on the double detection timescale}
    
    Considering the above uncertainties, we shall now derive a
    model-independent limit for the detection of double events, based
    on the fact that every UHE photon in a top-down model comes from
    the decay of a neutral pion, and is therefore accompanied by a
    second photon within a very small angle, due to relativistic
    beaming.  For a photon pair at $10^{20}$~eV, say, the parent pion
    Lorentz factor is $\Gamma_{\pi} = 2E/m_{\pi}c^{2} \simeq 1.4\times
    10^{12}$, so that the opening angle of this minimal, two-particle
    jet is $\theta_{\mathrm{jet}}\simeq 1/\Gamma_{\pi}\simeq 7\times
    10^{-13}$.  Using the source distance distribution,
    Eq.~(\ref{eq:p(D)}), and averaging over the decay angle in the
    pion rest frame, one finds the probability distribution of the
    distance $d$ between the two photons of a pair corresponding to a
    random UHECR event:
    \begin{equation}
	p(d) = \frac{\pi}{8}\frac{\Gamma_{\pi}}{D_{0}}\quad 
	\mathrm{for}\quad d\ll D_{0}/\Gamma_{\pi}.
    \end{equation}
    This allows us to estimate the probability that a detected UHE
    photon be accompanied by a second one within the range of the
    detector (of radius $R_{\mathrm{d}}$):
    \begin{equation}
	\mathcal{P}_{2} \simeq
	\frac{\pi}{16}\frac{\Gamma_{\pi}\left<R_{\mathrm{d}}\right>_{\perp}}
	{D_{0}}\simeq 7.8\times 10^{-5}
	\left(\frac{E}{10^{20}\,\mathrm{eV}}\right)
	\left(\frac{R_{\mathrm{d}}}{200\,\mathrm{km}}\right),
    \end{equation}
    From this, one can derive the minimum characteristic timescale for
    double event detection (independent of both
    $\theta_{\mathrm{jet}}$ and $N_{\gamma}$):
    \begin{equation}
	\tau_{2}^{\mathrm{min}} \simeq
	\frac{1}{\dot{N}_{\mathrm{evt}}\mathcal{P}_{2}} \simeq
	\frac{1}{\Phi^{\mathrm{td}}(\ge E)\pi^{2}R_{\mathrm{d}}^{2}
	\delta\, \mathcal{P}_{2}} \simeq 23\,\,\mathrm{yr}
	\left(\frac{\delta}{14\%}\right)
	\left(\frac{R_{\mathrm{d}}}{200\,\mathrm{km}}\right)^{-3}.
    \end{equation}
    While this timescale may seem prohibitively long (17 years for the
    EUSO detector, of radius $\sim 220$~km), one should note the cubic
    dependence in $R_{\mathrm{d}}$: a detector only two times larger
    (i.e. on an orbit two times higher) would detect double events
    from the \textit{minimal} top-down jets on a timescale slightly
    above 2 years.  This model-independent upper limit may give
    confidence that multiple event detection should indeed be
    possible, if the UHECRs are produced in Galactic top-down jets
    containing not just two photons from an isolated $\pi^{0}$ decay,
    but thousands of UHE photons.
    
    \section{Conclusion}

    In this paper, we have studied the possibility of observing
    multiple UHECR events from Galactic hadron jets, resulting from
    the decay of supermassive X-particles in the Halo.  We have shown
    that, under reasonable assumptions about the jet properties, the
    next generation UHECR detectors should be able to detect a few
    double events, and possibly one or two triple and quadruple
    events, provided the UHE flux is dominated by top-down sources at
    about $10^{20}$~eV, and the mass of the X-particle progenitor is
    around the GUT scale.

    The main uncertainties in our calculations come from the jet
    model.  Our assumptions about the photon spectrum, the jet
    multiplicity and the jet opening angle can only be considered as
    rough estimates, and other X-particle model may lead to different
    values.  Nevertheless, we have derived a general framework for the
    study of multiple events probability, and given a way to calculate
    the relevant detection timescales for any X-particle model, once
    its physical parameters are specified.  In particular, we have
    shown that the timescale $\tau_{2}$ is proportional to the jet
    opening angle, and inversely proportional to the square root of
    the photon multiplicity in the jets.

    Considering the jet uncertainties, we have also derived the double
    event detection timescale for the worst possible case: a
    two-particle jet consisting of the two photons produced by the
    decay of a neutral pion, as must be found in any Galactic hadron
    jet, whatever the model considered.  This gives an upper limit on
    $\tau_{2}$ which would reduce to $\sim 2$~years for a detector
    twice as large as EUSO. Note that this is also independent of the
    X-particle mass, contrary to the timescales obtained by taking
    into account all the particles inside the jet.  Note also that we
    have assumed homogeneously distributed photons inside the jets. 
    In the case of a clumpy distribution, the probability of observing
    a multiple event can only be higher than what we have obtained
    here, because the photon density close to an arbitrary photon
    would then be higher (on average) than the mean photon density in
    the jet.
    
    Besides this new test of Galactic top-down models, three other
    observational signatures have already been proposed.  First,
    top-down scenarios predict that photons should be the dominant
    component among UHECRs above $10^{20}$~eV. As we have seen, this
    is very important for our study, because charged particles of even
    ultra-high rigidity would be slightly deflected in the Galactic
    magnetic fields and lose the almost perfect collimation which
    multiple event require.  Second, the dipole anisotropy due to the
    off-centered position of the Earth should eventually show up in
    the data, although this indirect evidence might not be fully
    discriminatory, since at least one bottom-up model has been
    proposed with the same characteristics \cite{DarPla99}.  Finally,
    the UHECR energy spectrum should be characteristic of a hadronic
    fragmentation process, which may be quite different from the power
    laws usually expected from astrophysical acceleration processes. 
    However, the exact shape of the spectrum is still hard to predict
    precisely in any of the top-down or bottom-up models, and it is
    not clear whether a measurement with reasonable error bars over
    two decades in energy at most ($10^{19}$ to $10^{21}$~eV, say) can
    lead to definitive conclusions.
    
    By contrast, the detection of (were it only) one multiple event
    would provide a clear, direct evidence that a top-down scenario is
    involved, because it is highly improbable that two independent,
    bottom-up UHECRs arrive exactly at the same time from exactly the
    same direction.  In this respect, we should recall that a previous
    study excluded (or more precisely found very improbable) the
    possibility that a heavy nucleus be photodisintegrated by the
    solar radiation field and give rise to a pair of showers from the
    lighter, daughter nuclei \cite{PairsFromPhotoDis} (besides, such a
    pair would be perfectly correlated with the Sun's position).

    As shown above, the ideal detector to implement such a test and
    detect multiple events from Galactic-size hadron jets is a
    detector of very large acceptance, but not necessarily good
    angular and energy resolutions, which is usually one of the main
    challenges for UHECR detectors.  One might therefore think about
    the interest of devising a detector made of a series of
    atmospheric fluorescence telescopes covering a huge surface on
    Earth, but with poor angular and energy resolutions in order to
    keep it economical.
    
    Finally, we note that while the jet parameters are a major cause
    of uncertainty in the calculation of multiple events detection
    timescales, this very fact may offer a possibility to constrain
    them (if the top-down origin of UHECRs were to be attested by this
    or another way).  This may represent a unique opportunity to study
    the hadronization processes at energies many orders of magnitude
    above what can be reached in terrestrial accelerators.

    \section{Acknowledgements}
	
    I wish to thank warmly Yuri Dokshitzer and Michel Fontannaz for
    enlightening discussions and comments about QCD processes, jet
    formation and hadronization.

    \appendix
    \section{Appendix}
    
    In this appendix, we derive the general expression of the
    probability of detecting a multiple event as a function of the
    total number of UHECRs detected, which we shall note here $N$
    instead of $N_{evt}$, for convenience.  The probability of an
    event of multiplicity $m\ge n$ will be denoted by
    $\mathcal{P}_{\ge n}(N)$, and we shall prove that it can be
    written as in Eq.~(\ref{eq:globalProbaMultiN}):
    \begin{equation}
	\mathcal{P}_{\ge n}(N) = 1 - e^{-\frac{N}{N_{n}}},
        \label{eq:globalProbaMultiNApp}
    \end{equation}
    with the following values of the constants $N_{n}$:
    \begin{equation}
	N_{2} = \frac{1}{\sqrt\pi}\mu_{0}^{-1/2},\quad\mathrm{and}
	\quad N_{n+2} = \frac{2n}{2n-1}N_{n+1},
	\label{eq:N2App}
    \end{equation}
    which hold for small values of $n$, compared to the jet
    multiplicity $n\ll N_{\gamma}$.
    
    Each \textit{useful} X-particle decay event (i.e. giving rise to
    the detection of at least one UHECR), can be indexed by an integer
    $k$ ($1\le k\le N$), and is characterized by its distance,
    $D_{k}$, to the detector.  Its corresponding potential
    multiplicity is given by Eq.~(\ref{eq:muMu0}): $\mu_{k} =
    \mu_{0}D_{0}^{2}/D_{k}^{2}$.  The statistics of multiple events
    detection will therefore be determined by the statistics of the
    source distances, Eq.~(\ref{eq:p(D)}): $p(D) = 1/D_{0}$, for $0\le
    D\le D_{0}$.  Combining both expressions, we find the probability
    distribution of the random variable $\mu$:
    \begin{equation}
	p(\mu) = p(D)\left|\frac{\d D}{\d\mu}\right| =
	\frac{1}{2}\mu_{0}^{1/2}\mu^{-3/2}\quad\quad(\mu_{0}\le\mu\le\infty).
        \label{eq:p(mu)}
    \end{equation}
    
    We shall first establish that, for one particular (useful)
    X-particle decay event with potential multiplicity $\mu$, the
    probability that it is \textit{not} an event with actual
    multiplicity $m\ge n$ writes, for $n\ll N_{\gamma}$:
    \begin{equation}
	\tilde{\mathcal{P}}(m\ge n,\mu) = (1 +\mu +\frac{\mu^{2}}{2!}
	+ \ldots + \frac{\mu^{n-2}}{(n-2)!})\,e^{-\mu}.
        \label{eq:PTildeSupN}
    \end{equation}
    We proceed by recursion.  For $n=2$, the property reads
    $\tilde{\mathcal{P}}(m\ge 2,\mu) = 1 - \mathcal{P}(m\ge 2,\mu) =
    e^{-\mu}$, and has been obtained in Sect.~{\ref{sec:basicIdea}}. 
    Then:
    \begin{equation}
	\tilde{\mathcal{P}}(m\ge n+1,\mu) \equiv 1 - \mathcal{P}(m\ge
	n+1,\mu) = \tilde{\mathcal{P}}(m\ge n,\mu) + \mathcal{P}(n,\mu),
    \end{equation}
    where $\mathcal{P}(n,\mu)$ was given in
    Eq.~(\ref{eq:probaOfMulti}), and develops into:
    \begin{equation}
	\mathcal{P}(n,\mu) =
	\frac{\mu^{n-1}}{(n-1)!}
	\frac{(N_{\gamma}-1)\ldots(N_{\gamma}-(n-1))}
	{N_{\gamma}^{n-1}}
	\frac{\left(1-\frac{\mu}{N_{\gamma}}\right)^{N_{\gamma}}}
	{\left(1-\frac{\mu}{N_{\gamma}}\right)^{m}},
    \end{equation}
    and thus, for $n\ll N_{\gamma}$:
    \begin{equation}
	\mathcal{P}(n,\mu) \simeq \frac{\mu^{n-1}}{(n-1)!}e^{-\mu},
    \end{equation}
    which completes the proof.
    
    Now we consider the $N$ events together, and note that the
    `non-detection probabilities', $\tilde{\mathcal{P}}(m\ge n)$,
    simply multiply:
    \begin{equation}
        \tilde{\mathcal{P}}(m\ge n,\{\mu_{k}\}) = \prod_{k=1}^{N}
	\tilde{\mathcal{P}}(m\ge n,\mu_{k}).
    \end{equation}
    Now the global multiple event probability, $\mathcal{P}_{\ge
    n}(N)$, is the average value of $\mathcal{P}(m\ge n,\{\mu_{k}\}) =
    1 - \tilde{\mathcal{P}}(m\ge n,\{\mu_{k}\})$:
    \begin{equation}
	\begin{split}
	\mathcal{P}_{\ge n}(N) &= \left< 1 - \prod_{k=1}^{N}
	\tilde{\mathcal{P}}(m\ge n,\mu_{k}) \right> \\
	&= 1 - \left< \prod_{k=1}^{N} [(1 +\mu_{k} + \ldots +
	\frac{\mu_{k}^{n-2}}{(n-2)!})\,e^{-\mu_{k}}] \right>\\
	&= 1 - \left<\tilde{\mathcal{P}}(m\ge n,\mu)\right>^{N},
	\end{split}
    \end{equation}
    where the last equality follows from the statistical independence
    of the various $\mu_{k}$.
    
    This is indeed of the form announced in
    Eq.~(\ref{eq:globalProbaMultiNApp}), provided that we define the
    characteristic numbers $N_{n}$ by:
    \begin{equation}
	N_{n} \equiv \frac{-1}{\ln\left<\tilde{\mathcal{P}}(m\ge
	n,\mu)\right>}.
        \label{eq:NnApp}
    \end{equation}
    We thus have to calculate the average value of
    $\tilde{\mathcal{P}}(m\ge n,\mu)$:
    \begin{equation}
	\left<\tilde{\mathcal{P}}(m\ge n,\mu)\right>
	= \left<\tilde{\mathcal{P}}(m\ge n-1,\mu)\right> +
	\frac{1}{(n-2)!}\left<\mu^{n-2}e^{-\mu}\right>,
	\label{eq:PTildeRec}
    \end{equation}
    where, with the probability law of Eq.~(\ref{eq:p(mu)}):
    \begin{equation}
	\left<\mu^{n-2}e^{-\mu}\right> =
	\int_{\mu_{0}}^{\infty}\mu^{n-2}e^{-\mu}p(\mu)\d\mu =
	\frac{\mu_{0}^{1/2}}{2}
	\int_{\mu_{0}}^{\infty}\mu^{n-2-3/2}e^{-\mu}\d\mu.
    \end{equation}
    Integrating by parts, one finds:
    \begin{equation}
	\left<\mu^{n-2}e^{-\mu}\right> =
	\tfrac{1}{2}\mu_{0}^{n-3}e^{-\mu_{0}} + (n-2-\tfrac{3}{2})
	\left<\mu^{n-3}e^{-\mu}\right>.
    \end{equation}

    We shall now use the fact that $\mu_{0}\ll 1$ and limit the
    calculations to the lowest order (first order in $\mu_{0}^{1/2}$). 
    We can thus drop the first term in the right-hand side of the
    above equation, as long as $n\ge 4$, and rewrite it as:
    \begin{equation}
	\left<\mu^{n-2}e^{-\mu}\right> = (n-2-\tfrac{3}{2})
	\left<\mu^{n-3}e^{-\mu}\right>,\quad (n\ge 4),
    \end{equation}
    from where it follows, using Eq.~(\ref{eq:PTildeRec}) and writing
    $I_{n}\equiv\left<\tilde{\mathcal{P}}(m\ge n,\mu)\right>$ for
    simplicity, that
    \begin{equation}
	I_{n} - I_{n-1} = \frac{n-2-\tfrac{3}{2}}{n-2}(I_{n-1} -
	I_{n-2}),\quad (n\ge 4).
	\label{eq:InIn-1}
    \end{equation}
    This will allow us to calculate $I_{n}$ for all $n\ge 4$, once we
    know $I_{2}$ and $I_{3}$.  Starting with $I_{2} =
    \int_{\mu_{0}}^{\infty}e^{-\mu}p(\mu)\d\mu$ and integrating by
    parts, one finds:
    \begin{equation}
	I_{2} = e^{-\mu_{0}} - \mu_{0}^{1/2}
	\int_{\mu_{0}^{1/2}}^{\infty}\mu^{-1/2}e^{-\mu}\d\mu,
	\label{eq:step1}
    \end{equation}
    and then, by changing the variable to $u=\mu^{1/2}$:
    \begin{equation}
	I_{2} = e^{-\mu_{0}} -
	2\mu_{0}^{1/2}\int_{\mu_{0}^{1/2}}^{\infty}e^{-u^{2}}\d u
	\simeq e^{-\mu_{0}} - \sqrt\pi\mu_{0}^{1/2},
    \end{equation}
    where we have used $\mu_{0}\ll 1$ in the gaussian integral. 
    Developing to first order in $\mu_{0}^{1/2}$, we thus obtain:
    \begin{equation}
	I_{2} \simeq 1 - \sqrt\pi\mu_{0}^{1/2},
        \label{eq:I2Sol}
    \end{equation}
    and the first characteristic event number (for double event
    detection):
    \begin{equation}
	N_{2} = \frac{-1}{\ln I_{2}}\simeq
	\frac{1}{\sqrt\pi}\mu_{0}^{-1/2},
        \label{eq:N2Sol}
    \end{equation}
    as announced in (\ref{eq:N2App}).

    Coming now to $I_{3} =
    \int_{\mu_{0}}^{\infty}(1+\mu)e^{-\mu}p(\mu)\d\mu$, we have
    \begin{equation}
	I_{3} = I_{2} + \int_{\mu_{0}}^{\infty}\mu e^{-\mu}p(\mu)\d\mu
	= I_{2} + \frac{1}{2}\mu_{0}^{1/2}
	\int_{\mu_{0}^{1/2}}^{\infty}\mu^{-1/2}e^{-\mu}\d\mu,
    \end{equation}
    where we can replace the last term, using Eq.~(\ref{eq:step1}):
    \begin{equation}
        I_{3} = \frac{1}{2}(I_{2} + e^{-\mu_{0}})\simeq 1 - 
        \frac{\sqrt\pi\mu_{0}^{1/2}}{2}.
	\label{eq:I3Sol}
    \end{equation}
    This allows us to write the characteristic number of events for
    the detection of triple events, $N_{3} = -1/\ln I_{3}$, as:
    \begin{equation}
	N_{3} = \frac{2}{\sqrt\pi}\mu_{0}^{-1/2} = 2\times N_{2}.
        \label{eq:N3Sol}
    \end{equation}
    From (\ref{eq:I2Sol}) and~(\ref{eq:I3Sol}), we find
    \begin{equation}
        I_{3} - I_{2} = \frac{\sqrt\pi\mu_{0}^{1/2}}{2},
    \end{equation}
    which we can insert into (\ref{eq:InIn-1}) to obtain:
    \begin{equation}
        I_{4} = I_{3} + \frac{2-\frac{3}{2}}{2}(I_{3}-I_{2}) \simeq 
        1 - \frac{3}{8}\sqrt\pi\mu_{0}^{1/2},
    \end{equation}
    and thus
    \begin{equation}
	N_{4} = \frac{8}{3}\frac{\mu_{0}^{-1/2}}{\sqrt\pi} =
	\frac{4}{3}\times N_{3}.
        \label{eq:N4Sol}
    \end{equation}
    
    More generally, we can show that, at the first order in
    $\mu_{0}^{1/2}$,
    \begin{equation}
        I_{n} = 1 - \alpha_{n}\sqrt\pi\mu_{0}^{1/2}, 
        \quad\mathrm{with}\quad 
        \alpha_{n}=\alpha_{n-1}\frac{2(n-2)-1}{2(n-2)}.
	\label{InRecursion}
    \end{equation}
    We have already found that this is true for $n=2$ and $n=3$.  If
    we now assume that it is true for $n-2$ and $n-1$, we can use 
    Eq.~(\ref{eq:InIn-1}) to calculate $I_{n}$:
    \begin{equation}
	\begin{split}
	    I_{n} &= I_{n-1} + \frac{2(n-2)-3}{2(n-2)}(I_{n-1}-I_{n-2})\\
	    &= 1 - \sqrt\pi\mu_{0}^{1/2}\alpha_{n-1} -
	    \frac{2n-7}{2n-4}(\alpha_{n-1}-\alpha_{n-2})\sqrt\pi\mu_{0}^{1/2}\\
	    &= 1 - \sqrt\pi\mu_{0}^{1/2}\alpha_{n-1}\left[1 + \frac{2n-7}{2n-4}
	    - \frac{2n-7}{2n-4}\frac{2(n-3)}{2(n-3)-1}\right]\\
	    &= 1 - \sqrt\pi\mu_{0}^{1/2}\alpha_{n-1}\frac{2(n-2)}{2(n-2)-1},
	\end{split}
    \end{equation}
    which is indeed the recursion relation~(\ref{InRecursion}).
    
    The characteristic number of events for a multiplicity larger than
    $n$ is thus finally:
    \begin{equation}
        N_{n} = \frac{-1}{\ln I_{n}} = 
        \frac{1}{\alpha_{n}}\frac{\mu_{0}^{1/2}}{\sqrt\pi},
    \end{equation}
    and the recursion relation~(\ref{eq:N2App}) simply follows 
    from that in (\ref{InRecursion}). QED.
    
\end{document}